\documentclass{ws-mpla}

\begin{document}

\markboth{Ke-Sheng Sun,etc.}
{Lepton flavor violation decays of vector mesons in unparticle physics}

\catchline{}{}{}{}{}

\title{LEPTON FLAVOR VIOLATION DECAYS OF VECTOR MESONS IN UNPARTICLE PHYSICS}

\author{KE-SHENG SUN  $^{\dagger,\ddagger,\ast}$ ,
        TAI-FU FENG   $^{\ddagger,\dagger}$,
        LI-NA KOU     $^{\dagger}$ ,
        FEI SUN       $^{\dagger}$ ,\\
        TIE-JUN GAO   $^{\ddagger,\dagger}$,
        HAI-BIN ZHANG $^{\ddagger,\dagger}$
         }

\address{$^{\dagger}$    Department of Physics, Dalian University of Technology, Dalian 116024, China\\
         $^{\ddagger}$   Department of Physics, Hebei University, Baoding 071002,China\\
         $^{\ast}   $    sunkesheng@126.com }

\maketitle

\pub{Received (Day Month Year)}{Revised (Day Month Year)}

\begin{abstract}
We investigate the lepton flavor violation decays of vector mesons in the scenario
of the unparticle physics by considering the constraint from $\mu-e$ conversion. In unparticle physics, the predictions of LFV decays of vector mesons depend strongly on the scale dimension $d_{\mathcal{U}}$.
The predictions of LFV decays of vector mesons can reach the detective sensitivity in experiment in region of $3\le d_{\mathcal{U}}\le 4$, while the prediction of $\mu-e$ conversion rate can meet the experimental upper limit.
For the searching of the lepton flavor violation processes of charged lepton sector in experiment,
the process $\Upsilon\rightarrow e\mu$ may be a promising one to be observed.
\end{abstract}

\ccode{PACS Nos.:13.20.-v, 12.60.-i}

\section{Introduction}\label{intro}	

During the last decades, searching for Lepton Flavor Violation (LFV) processes in charged lepton sector, as an evidence to discover new physics beyond the Standard Model (SM), have attracted a great deal of attention.
Although nonzero neutrino masses supported by the neutrino oscillation experiments \cite{oscillation1,oscillation2,oscillation3} imply
the non-conservation of lepton flavor, due to the small masses of neutrinos, the lepton flavor violating processes in the SM are highly suppressed.
Nevertheless, the LFV decays could be enhanced by the new sources of LFV in various extensions of the SM, such as grand unified models,\cite{GUT1,GUT2,GUT3} supersymmetric models with and without R-parity,\cite{SUSY1,SUSY2,SUSY3} left-right symmetry models etc.\cite{LR1,LR2,LR3} These new sources are mainly originated  from the interactions between the SM particles and new particles beyond SM. Instead, Georgi proposes an alternative scenario that there could be a sector that is exactly scale invariant and very weakly interacting with the sectors in SM.\cite{Georgi1,Georgi2}
There are no particles in such a sector in space-times spaces cause no particles states with a nonzero mass exist. In general, the scale invariant sector or the so-called unparticle has a scale dimension of fractional number rather than an integral number.
The interactions between the unparticle and the SM particles in low energy effective theory can lead to various
interesting features in LFV processes and other phenomenologies.
In unparticle physics, the unparticle can interact with different flavors of SM leptons and this indicates that the LFV processes can happen at tree level. There have been many studies of LFV processes in unparticle physics. Such as, $\mu\rightarrow 3e$,\cite{Aliev,Choudhury} $\mu\rightarrow e\gamma$,\cite{Ding} $\mu-e$ conversion,\cite{Ding} $M^0\rightarrow ll'$,\cite{Lu} $e^+ e^-\rightarrow ll'$,\cite{Lu} $J/\Psi\rightarrow ll'$,\cite{Wei} $r\rightarrow ll'$,\cite{Iltan1,Iltan2} $l\rightarrow l'\gamma\gamma$,\cite{Iltan1,Iltan2} $\tau\rightarrow l(V_0,P_0)$,\cite{Li} etc..

The study of LFV processes involving vector mesons is an effective way maybe to search for new physics beyond the SM, and the SND Collaboration at the BINP (Novosibirk) presents an upper limit on the $\phi\rightarrow e^+\mu^-$ branching fraction of BR$(\phi\rightarrow e^+\mu^-)\le 2\times 10^{-6}$ .\cite{SND}
Additionally, using a sample of $5.8\times10^7\;J/\Psi$ events collected with the BESII detector,
Ref.~\refcite{BESII} obtains the upper limits on BR$(J/\Psi\rightarrow\mu\tau)<2.0\times10^{-6}$
and BR$(\Upsilon\rightarrow\mu\tau)<8.3\times10^{-6}$ at the 90\% confidence level (C.L.).
Adopting the data collected with the CLEO III detector, the authors of Ref.~\refcite{CLEO} estimate the upper limits on BR $(\Upsilon(1S)\rightarrow\mu\tau)<6.0\times10^{-6}$, BR $(\Upsilon(2S)\rightarrow\mu\tau)<1.4\times10^{-5}$ and BR $(\Upsilon(3S)\rightarrow\mu\tau)<2.0\times10^{-5}$ respectively at the  95\% C.L.
In literatures, several stringent limits on LFV decays of vector mesons are derived in a model independent way. Assuming that a vector boson $M_i$ couples to $\mu^{\mp}e^{\pm}$ and $e^{\mp}e^{\pm}$, the authors of Ref.~\refcite{zhang} deduce some upper bounds on the LFV decays of vector mesons by a consideration of the experimental constraint on the process $\mu\rightarrow3e$.
Under a similar assumption that a vector meson $M_i$ couples to $\mu^{\mp}e^{\pm}$ and nucleon-nucleon, Ref.~\refcite{Gutsche1} and Ref.~\refcite{Gutsche2} study the LFV decays of vector mesons by taking account of the experimental constraint on $\mu-e$ conversion.

In this paper, we investigate the LFV decays of vector mesons in unparticle physics by the consideration of constraint on $\mu-e$ conversion. In Section.\ref{sec:2}, we firstly provide a brief introduction to the unparticle physics and corresponding interaction Lagrangian in effective field theory. Then we derive the analytic results of the amplitude in detail. The numerical analysis and discussion are presented in Section.\ref{sec:3}, and the conclusion is drawn in Section.\ref{sec:4}.

\section{Formalism}
\label{sec:2}

In very high energy, as it is proposed by Georgi,\cite{Georgi1,Georgi2} the theory is composed of the SM fields and the fields of a theory with a nontrivial IR fixed point, which is called Banks-Zaks ($\mathcal{BZ}$) fields.\cite{BZ} The two fields can interact by the exchange of particles with a large mass scale $M_{\mathcal{U}}\gg1TeV$. Below the scale $M_{\mathcal{U}}$, there are nonrenormalizable couplings involving both standard model fields and Banks-Zaks fields suppressed by powers of $M_{\mathcal{U}}$. The interaction between SM field and $\mathcal{BZ}$ field has the form:
\begin{eqnarray}
\frac{1}{M_{\mathcal{U}}^{d_{SM}+d_{\mathcal{BZ}}-4}}O_{SM}O_{\mathcal{BZ}},
\label{SM-BZ}
\end{eqnarray}
where $O_{SM}$ is an operator with a mass dimension $d_{SM}$ corresponding to SM fields and $O_{\mathcal{BZ}}$ is an operator with a mass dimension $d_{\mathcal{BZ}}$ corresponding to $\mathcal {BZ}$ fields.
In effective field theory, below the scale $\Lambda_{\mathcal{U}}$, the $\mathcal {BZ}$ operators would match onto the unparticle operators, and Eq.(\ref{SM-BZ}) can be viewed as the interaction between SM field and unparticle field:
\begin{eqnarray}
\frac{C_{\mathcal{U}}\Lambda^{d_{\mathcal{BZ}}-d_{\mathcal{U}}}_{\mathcal{U}}}
{M_{\mathcal{U}}^{d_{SM}+d_{\mathcal{BZ}}-4}}O_{SM}O_{\mathcal{U}},
\label{SM-U}
\end{eqnarray}
where $C_{\mathcal{U}}$ is a coefficient function, $d_{\mathcal{U}}$ denotes the scaling dimension of the unparticle operator $O_{\mathcal{U}}$.

For simplicity, it is convenient to define:
\begin{eqnarray}
\lambda=\frac{C_{\mathcal{U}}\Lambda^{d_{\mathcal{BZ}}}_{\mathcal{U}}}{M_{\mathcal{U}}^{d_{SM}+d_{\mathcal{BZ}}-4}}.
\label{Lam}
\end{eqnarray}
Then, in effective theory, the couplings of the scalar and vector unparticles to SM fermions (leptons or quarks) are generally given by the following effective operators:
\begin{eqnarray}
&&\frac{\lambda^{SS}_{ff'}}{\Lambda^{d_{\mathcal{U}}-1}_{\mathcal{U}}}\bar{f}f'O_{\mathcal{U}},
\frac{\lambda^{SP}_{ff'}}{\Lambda^{d_{\mathcal{U}}-1}_{\mathcal{U}}}\bar{f}\gamma_{5}f'O_{\mathcal{U}},
\frac{\lambda^{SV}_{ff'}}{\Lambda^{d_{\mathcal{U}}}_{\mathcal{U}}}\bar{f}\gamma_{\mu}f'\partial^{\mu}{O_{\mathcal{U}}},
\frac{\lambda^{SA}_{ff'}}{\Lambda^{d_{\mathcal{U}}}_{\mathcal{U}}}\bar{f}\gamma_{\mu}\gamma_{5}f'\partial^{\mu}{O_{\mathcal{U}}},
\nonumber\\
&&\frac{\lambda^{VS}_{ff'}}{\Lambda^{d_{\mathcal{U}}}_{\mathcal{U}}}\bar{f}f'\partial_{\mu}{O^{\mu}_{\mathcal{U}}},
\frac{\lambda^{VP}_{ff'}}{\Lambda^{d_{\mathcal{U}}}_{\mathcal{U}}}\bar{f}\gamma_{5}f'\partial_{\mu}{O^{\mu}_{\mathcal{U}}},
\frac{\lambda^{VV}_{ff'}}{\Lambda^{d_{\mathcal{U}}-1}_{\mathcal{U}}}\bar{f}\gamma_{\mu}f'O^{\mu}_{\mathcal{U}},
\frac{\lambda^{VA}_{ff'}}{\Lambda^{d_{\mathcal{U}}-1}_{\mathcal{U}}}\bar{f}\gamma_{\mu}\gamma_{5}f'O^{\mu}_{\mathcal{U}},
\label{Eff-ope}
\end{eqnarray}
where $\lambda^{S,P,V,A}_{ff'}$ are dimensionless coefficients. $S,P,V$ and $A$ stand for scalar field, pseudo-scalar field, vector field and axial vector field, respectively.$f$ and $f'$ denote SM fermions, $O_{\mathcal{U}}$ and $O^{\mu}_{\mathcal{U}}$ denote scalar and vector unparticle fields.
The propagator of scalar unparticle field has the form\cite{Georgi2,Cheung}:
\begin{eqnarray}
\int&& e^{iP\cdot x}d^4 x \langle 0|T[O_{\mathcal{U}}(x)O_{\mathcal{U}}(0)|0\rangle = i\frac{A_{d_{\mathcal{U}}}}{2 \sin(d_{\mathcal{U}}\pi)}\frac{1}{(-P^2-i\epsilon)^{d_{\mathcal{U}}-2}}
\end{eqnarray}
If the vector unparticle field is assumed to be transverse, the propagator can been written as:
\begin{eqnarray}
\int&& e^{iP\cdot x}d^4 x \langle 0|T[O^{\mu}_{\mathcal{U}}(x)O^{\nu}_{\mathcal{U}}(0)|0\rangle = i\frac{A_{d_{\mathcal{U}}}}{2 \sin(d_{\mathcal{U}}\pi)}\frac{-g^{\mu\nu}+P^{\mu}P^{\nu}}{(-P^2-i\epsilon)^{2-d_{\mathcal{U}}}}
\end{eqnarray}
where $A_{d_{\mathcal{U}}}$ is defined by:
\begin{eqnarray}
A_{d_{\mathcal{U}}}=\frac{16\pi^{5/2}}{(2\pi)^{2d_{\mathcal{U}}}}
\frac{\Gamma(d_{\mathcal{U}}+1/2)}{\Gamma(d_{\mathcal{U}}-1)\Gamma(2d_{\mathcal{U}})}
\end{eqnarray}

\begin{figure}[h]
\centerline{\psfig{file=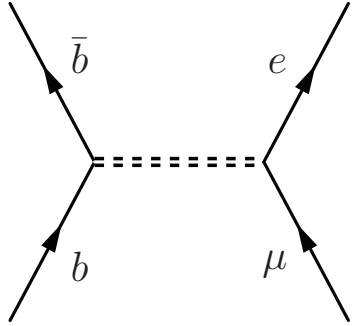,width=2.0in}}
\vspace*{8pt}
\caption{The tree level diagram for LFV process $\Upsilon\rightarrow e\mu$, the double dashed line denotes the vector unparticle field.\protect\label{fig1}}
\end{figure}

For vector mesons, only the vector current $\bar{f}\gamma_{\mu}f'$ couples to vector mesons.
The tree level Feynman diagram is presented in Fig.\ref{fig1}.
The amplitude for Fig.\ref{fig1} can be written as:
\begin{eqnarray}
{\cal M}_Q=&& \frac{\lambda^{VV}_{bb}\lambda^{VV}_{e \mu}}{\Lambda^{2d_{\mathcal{U}}-2}_{\mathcal{U}}}
\frac{A_{d_{\mathcal{U}}}}{2\sin(d_{\mathcal{U}}\pi)}\bar{\upsilon}(p_1)\gamma_{\mu}u(p_2)
\frac{g^{\mu\nu}-p^{\mu}p^{\nu}}{p^{2(2-d_{\mathcal{U}})}}\bar{u}(p_3)\gamma_{\nu}\upsilon(p_4).
\label{Amp}
\end{eqnarray}
In the quark picture, mesons are composed of a quark and an antiquark. We adopt a phenomenological model
where the amplitude of hard process involving a s-wave meson
can be described by the matrix elements of gauge-invariant nonlocal operators,
which are sandwiched between the vacuum and the meson states. The distribution amplitude of vector meson $\Upsilon$ in leading-order is defined through the correlation function\cite{OZI1,OZI2,OZI3}:
\begin{eqnarray}
\langle 0| \bar{b}_{1\alpha }^{i}(y)b_{2\beta}^{j}(x)|\Upsilon(p)\rangle
&=&\frac{\delta_{ij}}{4N_{c}}\int_{0}^{1}du~e^{-i[upy+(1-u)px]}
\Big[ f_{\Upsilon }m_{\Upsilon }/\!\!\!\varepsilon \phi _{\parallel}(u)
\nonumber\\
&&+\frac{i}{2}\sigma ^{{\mu }'{\nu }'}f_{\Upsilon }^{T}
( \varepsilon _{ {\mu }'}{p}_{{\nu }'}-\varepsilon _{{\nu }'}{p}_{{\mu }'}
)\phi _{\perp } ( u )\Big]_{\beta \alpha}\;,
\label{hadron}
\end{eqnarray}
where Nc is the number of colors, $\varepsilon$ is the polarization vector, $f_{\Upsilon }$ and $f_{\Upsilon}^{T}$ are the decay constants, $\phi _{\parallel }$ and $\phi _{\perp }$ are the
leading-twist distribution functions corresponding to the longitudinally
and transversely polarized meson, respectively. Since the leading-twist light-cone
distribution amplitudes of meson are close to their asymptotic form,\cite{Beneke}
so we set $\phi _{\parallel }=\phi _{\perp }=\phi(u)=6u(1-u)$.

Then, at hadron level, using Eq.(\ref{hadron}), the amplitude is rewritten as
\begin{eqnarray}
{\cal M}_H= \frac{\lambda^{VV}_{bb}\lambda^{VV}_{e \mu}}{\Lambda^{2d_{\mathcal{U}}-2}_{\mathcal{U}}}
\frac{A_{d_{\mathcal{U}}}m_{\Upsilon}f_{\Upsilon}}{2N_c\sin(d_{\mathcal{U}}\pi)}\frac{\varepsilon^{\nu}}
{p^{2(2-d_{\mathcal{U}})}}\bar{u}(p_3)\gamma_{\nu}\upsilon(p_4).
\end{eqnarray}
In the frame of center of mass, using the summation formula
\begin{eqnarray}
\sum_{\lambda =\pm 1,0}\varepsilon ^{\mu }_{\lambda }(p)\varepsilon ^{\ast \nu }_{\lambda }(p)
\equiv -g^{\mu \nu }+\frac{p^{\mu }p^{\nu }}{m_{\Upsilon}^{2}},
\end{eqnarray}
we get
\begin{eqnarray}
|{\cal M}_H|^2 = \mid\frac{\lambda^{VV}_{bb}\lambda^{VV}_{e \mu}}{\Lambda^{2d_{\mathcal{U}}-2}_{\mathcal{U}}}
\frac{A_{d_{\mathcal{U}}}m_{\Upsilon}f_{\Upsilon}}{2N_c\sin(d_{\mathcal{U}}\pi)}\mid^2
\frac{4(m^2_{\Upsilon}-m^2_{e}-m^2_{\mu})-16 m_e m_{\mu}}{m_{\Upsilon}^{4(2-d_{\mathcal{U}})}}.
\label{MH}
\end{eqnarray}
Finally,we express the branching ratio of process $\Upsilon\rightarrow e\mu$ as
\begin{eqnarray}
Br(\Upsilon\rightarrow e\mu)=\frac{\sqrt{ [m_{\Upsilon}^{2}-(m_{e}+m_{\mu})^{2}]
[m_{\Upsilon}^{2}-(m_{e}-m_{\mu})^{2}] }}{16 \pi m_{\Upsilon}^{3}\Gamma_{\Upsilon}}\times|{\cal M}_H|^2,
\label{BR}
\end{eqnarray}
where $\Gamma_{\Upsilon}$ is the total decay width. The branching ratios for
other LFV processes of vector mesons can be formulated in a similar way.

\section{Numerical Analysis}
\label{sec:3}

Taking account of the constraint on the LFV processes
$\mu\rightarrow e\gamma$ and $\mu-e$ conversion in nuclei, we will study the
LFV decay of $\Upsilon\rightarrow e\mu$ in unparticle physics firstly. Under the assumption of
the unparticle couplings with the SM fermions in Eq.(\ref{Eff-ope}) are universal:
\begin{equation}\label{Assume}
\lambda^{KK}_{ff'} =
\left\{
\begin{aligned}
\lambda_{k},  & f=f' \\
\kappa\lambda_{k}, & f\neq f'
\end{aligned}
\right.
\end{equation}
where $\kappa > 1$ and K = S, P, V, A for scalar, pseudoscalar, vector and axial
vector couplings respectively, the authors of Ref.~\refcite{Ding} have investigated the LFV processes $\mu\rightarrow e\gamma$
and $\mu-e$ conversion in various nuclei in region of $1<d_{\mathcal{U}}<2$,
$1 \mathrm{TeV}<\Lambda_{\mathcal{U}}<100 \mathrm{TeV}$. It displays the constraint on dimension $d_{\mathcal{U}}$ deduced from experimental bound on $\mu-e$ conversion is more stringent. Therefore, we will study the
LFV decays of vector mesons by a consideration of $\mu-e$ conversion in unparticle physics. The formula for the $\mu-e$ conversion rate with the pure vector coupling between SM fermions and unparticle is given by:
\begin{eqnarray}
CR(\mu-e, Nucleus)&=&\frac{m^5_{\mu} \alpha^3 Z^4_{eff} F^2_p}{2 \pi^2 Z }
[\lambda^{VV}_{e\mu}\lambda^{VV}_{qq}\frac{A_{d_{\mathcal{U}}}}{2\sin(d_{\mathcal{U}}\pi)}
\frac{1}{\Lambda^2_{\mathcal{U}}}(\frac{m^2_{\mu}}{\Lambda^2_{\mathcal{U}}})^{d_{\mathcal{U}}-2}]^2
\nonumber\\
&&\times\mid Z\sum_q G^{(q,p)}_V + N \sum_q G^{(q,n)}_V\mid^2\frac{1}{\Gamma_{capt}},
\label{CR}
\end{eqnarray}
where Z and N denote the proton and neutron numbers in a nucleus, $F_p$ is the
nuclear form factor and $Z_{eff}$ is an effective atomic charge, $G^{(q,p)}_V$
and $G^{(q,n)}_V$ are nuclear matrix elements relevant to proton and neutron.

\begin{figure}[h]
\centerline{\psfig{file=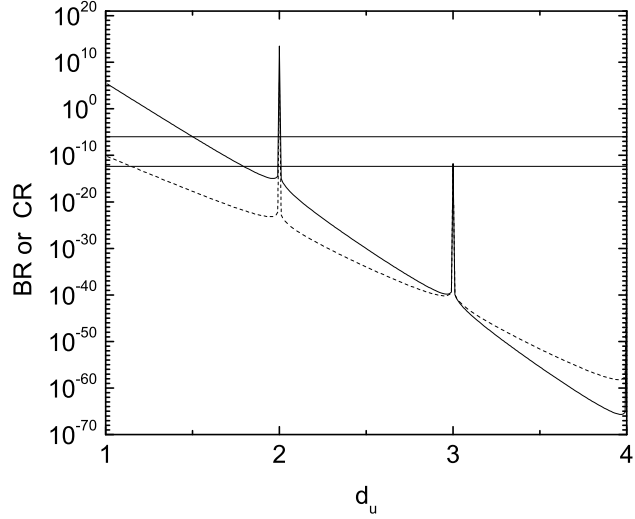,width=4.0in}}
\vspace*{8pt}
\caption{The BR$(\Upsilon\rightarrow e\mu)$ and CR$(\mu-e, Ti)$ vary as a function of $d_{\mathcal{U}}$, where the solid line denotes the prediction of CR$(\mu-e, Ti)$, the dot line denotes the prediction of BR$(\Upsilon\rightarrow e\mu)$, the horizontal lines correspond to $10^{-6}$ and $4.2\times10^{-12}$, which are the experimental sensitivity of LFV decays of vector mesons and the experimental bound on $\mu-e$ conversion rate respectively. $\Lambda_{\mathcal{U}}=10$TeV, $\lambda^{VV}_{bb}=0.001$ and $\lambda^{VV}_{e\mu}=0.003$ are assumed.\protect\label{fig2}}
\end{figure}

Here, as in Ref.~\refcite{Ding}, taking $\Lambda_{\mathcal{U}}=10$TeV,
$\lambda^{VV}_{bb}=0.001$ and $\lambda^{VV}_{e\mu}=0.003$, we display BR$(\Upsilon\rightarrow e\mu)$ versus $d_{\mathcal{U}}$ and CR$(\mu-e, Ti)$ versus $d_{\mathcal{U}}$ in region of $1\le d_{\mathcal{U}}\le 4$ in Fig.\ref{fig2},
where the solid line denotes the prediction of CR$(\mu-e, Ti)$, the dot line denotes the prediction of BR$(\Upsilon\rightarrow e\mu)$. The horizontal lines correspond to $10^{-6}$ and $4.2\times10^{-12}$, which are the experimental sensitivity of LFV decays of vector mesons and the experimental bound on $\mu-e$ conversion rate respectively.
The following numerical values are used\cite{Decay constant1,Decay constant2,Decay constant3}:
\begin{eqnarray}
&&m_{\Upsilon}=9.406GeV,f_{\Upsilon}=715MeV,\Gamma_{\Upsilon}=54KeV,\nonumber\\
&&F_p=0.54,Z_{eff}=17.6, \Gamma_{capt}=1.7\times10^{-18}.
\end{eqnarray}

As we can see from Fig.\ref{fig2}, the predictions of both BR$(\Upsilon\rightarrow e\mu)$ and CR$(\mu-e, Ti)$
depend strongly on the scaling dimension $d_{\mathcal{U}}$. The value of dimension $d_{\mathcal{U}}$
is constrained to near 2 or more larger and the relevant prediction of BR$(\Upsilon\rightarrow e\mu)$ is highly suppressed to reach the experimental sensitivity.

There is an interesting feature in Fig.\ref{fig2} that the prediction of
BR$(\Upsilon\rightarrow e\mu)$ is less than CR$(\mu-e, Ti)$ in region of $1\le d_{\mathcal{U}}\le 3$, however, in region of $3\le d_{\mathcal{U}}\le 4$, the prediction of BR$(\Upsilon\rightarrow e\mu)$ is larger than CR$(\mu-e, Ti)$. Considering the experimental bound on CR$(\mu-e, Ti)$ is $\mathcal{O} (10^{-12})$, it is impossible to make the prediction of BR$(\Upsilon\rightarrow e\mu)$ reach the experimental sensitivity by resetting the couplings $\mid\lambda^{VV}_{bb}\lambda^{VV}_{e\mu}\mid$ in region of $1\le d_{\mathcal{U}}\le 3$. Nevertheless, in region of $3\le d_{\mathcal{U}}\le 4$, by enlarging the couplings $\mid\lambda^{VV}_{bb}\lambda^{VV}_{e\mu}\mid$, it is available to get that not only the prediction of CR$(\mu-e, Ti)$ is compatible with the experimental bound, but also the prediction of BR$(\Upsilon\rightarrow e\mu)$ is large enough to be detected in experiment at present or in near future. Therefore, we will investigate the LFV decays of vector mesons in region of $3\le d_{\mathcal{U}}\le 4$. In addition, the constraint of $d_{\mathcal{U}}\geq3$ is also supported in Ref.~\refcite{Mark1} and Ref.~\refcite{Mark2} by the consideration of unitarity. From unitarity, the gauge invariant primary vector operator $\mathcal{U}^{\mu}$ have $d_V\ge3$, with $d_V=3$ if and only if the operator is a conserved current, $\partial_{\mu}\mathcal{U}^{\mu}=0$.

For the aim of enhancing the prediction of BR$(\Upsilon\rightarrow e\mu)$ to be detectable in experiment, the value of the couplings $\mid\lambda^{VV}_{bb}\lambda^{VV}_{e\mu}\mid$ would be very large. However, we can also investigate the LFV decays of vector mesons in a way independent of the couplings $\mid\lambda^{VV}_{bb}\lambda^{VV}_{e\mu}\mid$.
Considering the $\mu-e$ conversion in $Ti$ nucleus, let us define the fraction $R(X)$ by:
\begin{eqnarray}
R(X)=\frac{BR(X\rightarrow e\mu)}{CR(\mu-e,Ti)},
\label{Rx}
\end{eqnarray}
where $X$ would be any vector mesons:$\rho$, $\omega$, $\phi$, $J/\Psi$ or $\Upsilon$. From Eq.(\ref{MH}), Eq.(\ref{BR}) and Eq.(\ref{CR}), we can see that coefficient $\lambda^{VV}_{e\mu}$ and mass scale $\Lambda_{\mathcal{U}}$ can be canceled out in $R(X)$. As for the unparticle couplings with the quarks are universal, i.e.,
\begin{eqnarray}
\lambda^{VV}_{bb}=\lambda^{VV}_{ss}=\lambda^{VV}_{cc}\simeq
\frac{\lambda^{VV}_{uu}+\lambda^{VV}_{dd}}{\sqrt{2}}\simeq
\frac{\lambda^{VV}_{uu}-\lambda^{VV}_{dd}}{\sqrt{2}},
\label{assume}
\end{eqnarray}
the couplings listed in Eq.(\ref{assume}) would also be canceled out in $R(X)$ due to the same value setting in both numerator and denominator in Eq.(\ref{Rx}). Therefore, $R(X)$ would only be a function of scaling dimension $d_{\mathcal{U}}$. Using Eq.(\ref{MH}), Eq.(\ref{BR}) and Eq.(\ref{CR}), the relation between $R(X)$ and $d_{\mathcal{U}}$ can be shown in a simple form:
\begin{eqnarray}
R(X)\propto(\frac{m_X}{m_{\mu}})^{4(d_{\mathcal{U}}-2)},
\label{Relation}
\end{eqnarray}
where $\frac{m_X}{m_{\mu}}>1$ for different mesons. It is noteworthy that even if the unparticle
couplings with the SM fermions are not universal, i.e., Eq.(\ref{assume}) is not feasible, the relation between $R(X)$ and $d_{\mathcal{U}}$ in Eq.(\ref{Relation}) is still reliable.

\begin{figure}[h]
\centerline{\psfig{file=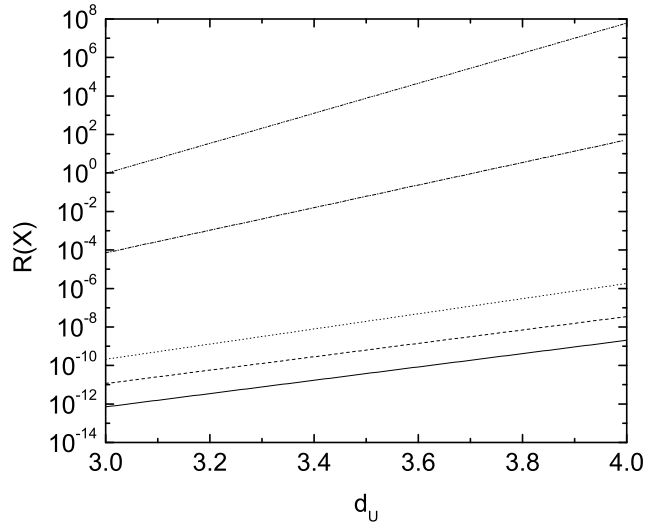,width=4.0in}}
\vspace*{8pt}
\caption{The fraction $R(X)$ varies as a function of $d_{\mathcal{U}}$, where, from the bottom up, the lines stand for $R(\rho)$,$R(\omega)$,$R(\phi)$,$R(J/\Psi)$ and $R(\Upsilon)$, respectively.\protect\label{fig3}}
\end{figure}

In Fig.\ref{fig3}, we display the fraction $R(X)$ varies as a function of $d_{\mathcal{U}}$ in region of $3\le d_{\mathcal{U}}\le 4$, where, from the bottom up, the lines stand for $R(\rho)$,$R(\omega)$,$R(\phi)$,$R(J/\Psi)$ and $R(\Upsilon)$, respectively. The parameters relevant to different mesons are listed below\cite{Decay constant1,Decay constant2,Decay constant3}:
\begin{eqnarray}
m_{\rho}&=&775MeV,f_{\rho}=209MeV,\Gamma_{\rho}=149MeV,\nonumber\\
m_{\omega}&=&782MeV,f_{\omega}=195MeV,\Gamma_{\omega}=8.49MeV,\nonumber\\
m_{\phi}&=&1.019GeV,f_{\phi}=231MeV,\Gamma_{\phi}=4.2MeV,\nonumber\\
m_{J/\Psi}&=&3.096GeV,f_{J/\Psi}=405MeV,\Gamma_{J/\Psi}=92.9KeV.\nonumber
\end{eqnarray}
It displays in Fig.\ref{fig3} that $R(X)$ increases as $d_{\mathcal{U}}$ grows. For light flavor mesons, the fraction $R(X)$ is very small. However, for heavy flavor mesons, the fraction $R(X)$ is large and we can get a large BR$(X\rightarrow e\mu)$ to be detectable in experiment and compatible with the constraint on $\mu-e$ conversion.

From Eq.(\ref{Rx}) we can express the branching ratio of $X\rightarrow e\mu$ as:
\begin{eqnarray}
BR(X\rightarrow e\mu)=R(X)\times CR(\mu-e,Ti).
\label{BRR}
\end{eqnarray}
Using Eq.(\ref{BRR}), we give the predictions on branching ratios of LFV decays of vector mesons in Tab.\ref{tab1} with $d_{\mathcal{U}}=3$, $d_{\mathcal{U}}=3.5$ and $d_{\mathcal{U}}=4$, where CR$(\mu-e,Ti)\le 4.2\times10^{-12}$ is used.
\begin{table}[h]
\tbl{Predictions on branching ratios of LFV decays of vector mesons with $d_{\mathcal{U}}=3$, $d_{\mathcal{U}}=3.5$ and $d_{\mathcal{U}}=4$, where CR$(\mu-e,Ti)\le 4.2\times10^{-12}$ is used.}
{\begin{tabular}{@{}cccc@{}} \toprule
Decay &$d_{\mathcal{U}}$=3&$d_{\mathcal{U}}$=3.5&$d_{\mathcal{U}}$=4\\
\colrule
$\rho\rightarrow e\mu$&$2.8\times10^{-24}$&$1.5\times10^{-22}$&$8.1\times10^{-21}$\\
$\omega\rightarrow e\mu$&$4.6\times10^{-23}$&$2.5\times10^{-21}$&$1.3\times10^{-19}$\\
$\phi\rightarrow e\mu$&$8.4\times10^{-22}$&$7.8\times10^{-20}$&$7.2\times10^{-18}$\\
$J/\Psi\rightarrow e\mu$&$2.8\times10^{-16}$&$2.4\times10^{-13}$&$2.1\times10^{-10}$\\
$\Upsilon\rightarrow e\mu$&$3.8\times10^{-12}$&$3.0\times10^{-8}$&$2.4\times10^{-4}$\\ \botrule
\end{tabular}\label{tab1} }
\end{table}
For light flavor mesons, the prediction of BR$(X\rightarrow e\mu)$ is much little, and it is impossible to observe the LFV processes of these mesons in experiment. For heavy flavor mesons, the large prediction of BR$(X\rightarrow e\mu)$ is available for $d_{\mathcal{U}}$ near 4.
Especially, the prediction of BR$(\Upsilon\rightarrow e\mu)$ is as large as $\mathcal{O}(10^{-4})$, and it is very  promising to be observed in experiment.

In literatures, several stringent limits on LFV decays of vector mesons are derived already. A summary table of  experimental bounds and corresponding theoretical predictions is presented in Tab.\ref{tab2}.
\begin{table}[h]
\tbl{The upper bounds on the branching ratios of vector meson decays in experiment and literatures.}
{\begin{tabular}{@{}cccc@{}} \toprule
Decay & Exp & Ref.~\refcite{zhang} &Ref.~\refcite{Gutsche}\\
\colrule
$\rho\rightarrow e\mu$&$-$&$\le3.8\times10^{-20}$&$\le3.5\times10^{-24}$\\
$\omega\rightarrow e\mu$&$-$&$\le8.1\times10^{-16}$&$\le6.2\times10^{-27}$\\
$\phi\rightarrow e\mu$&$\le2.0\times 10^{-6}$&$\le4.0\times10^{-17}$&$\le1.3\times10^{-21}$\\
$J/\Psi\rightarrow e\mu$&$<1.1\times10^{-6}$&$\le4.0\times10^{-13}$&$\le3.5\times10^{-13}$\\
$\Upsilon\rightarrow e\mu$&$-$&$\le2.0\times10^{-9}$&$\le3.9\times10^{-6}$\\ \botrule
\end{tabular}\label{tab2} }
\end{table}
Assuming that a vector boson $M_i$ couples to $\mu^{\mp}e^{\pm}$ and $e^{\mp}e^{\pm}$, the authors of Ref.~\refcite{zhang} deduce some upper bounds on the LFV decay of mesons using the experimental constraint on the LFV process $\mu\rightarrow3e$.
Under a similar assumption that a vector meson $M_i$ couples to $\mu^{\mp}e^{\pm}$ and nucleon-nucleon, Ref.~\refcite{Gutsche1} and Ref.~\refcite{Gutsche2} study the LFV decays of vector mesons by taking account of the experimental constraint on $\mu-e$ conversion.
From Tab.\ref{tab1} and Tab.\ref{tab2}, it is easy to find that our predictions are compatible with those in literatures.

Finally, the predictions of LFV decays of vector mesons in both our article and Ref.~\refcite{zhang,Gutsche1,Gutsche2} greatly depend on the experimental constraints on BR$(l_i\rightarrow l_j\gamma)$, BR$(l_i\rightarrow 3l_j)$ and CR$(\mu-e)$. Thus, more reliable predictions on LFV decays of vector mesons depend on the new data from the experiment. In the future, the expected sensitivities for BR $(\mu\rightarrow e\gamma)$ would be of order $10^{-13}$.\cite{MEG} For BR $(\tau\rightarrow e\gamma)$ and BR $(\tau\rightarrow\mu\gamma)$, it would be $10^{-9}$.\cite{Bona} For CR $(\mu-e, Ti)$, it would be as low as $10^{-16}\sim10^{-17}$.\cite{AIP}
Then, the predictions of BR$(X\rightarrow e\mu)$ for vector mesons would be more stringent.

\section{Conclusions\label{sec:4}}

Considering the constraint on $\mu-e$ conversion, we analyze the LFV decays of vector mesons in the framework of unparticle physics.
In the scenario of the unparticle physics, the predictions of branching ratios of LFV decays of vector mesons depend strongly on the scale dimension $d_{\mathcal{U}}$.
Supposing the unparticle couplings with the SM fermions are universal, the predictions of the branching ratios of the LFV decays of vector mesons can reach the detective sensitivity in experiment in region of $d_{\mathcal{U}}\geq3$, while the prediction of $\mu-e$ conversion rate in Ti nucleus is compatible with the experimental upper limit.
Although nonzero neutrino masses supported by the neutrino oscillation experiments imply the nonconservation of lepton flavors, it is very important to directly search the LFV processes of charged lepton sector in colliders running now.
The LFV process $\Upsilon\rightarrow e\mu$ is very promising to be observed in experiment.

\section*{Acknowledgements}

The work has been supported by the National Natural Science Foundation of China (NNSFC)
with Grants No. 10975027.

\end{document}